\documentclass[12pt]{article}
\usepackage{amssymb,amsmath}
\usepackage[noblocks]{authblk}
\usepackage[top=0.75in, bottom=0.75in, left=0.75in, right=0.75in, dvips]{geometry}
\usepackage{caption}
\date{}

\begin{document}
\textwidth 10.0in 
\textheight 9.0in 
\topmargin -0.60in
\title{Canonical Analysis of a System with Fermionic Gauge Symmetry}
\author[1,2]{D.G.C. McKeon}
\affil[1] {Department of Applied Mathematics, The
University of Western Ontario, London, ON N6A 5B7, Canada}
\affil[2] {Department of Mathematics and
Computer Science, Algoma University, Sault St.Marie, ON P6A
2G4, Canada}
\maketitle

\maketitle

\begin{abstract}
A non-Abelian gauge field with a topological action is coupled to a spin $3/2$ Majorana spinor. The symmetries of this model are analyzed using the Dirac constraint formalism.  These symmetries include a Fermionic symmetry and the algebra of these symmetries closes; it is not the algebra of supergravity. The action is invariant without the need to introduce auxiliary fields.
\end{abstract}

\section{Introduction}
The idea of extending the Poincare group through use of anti-commuting (Fermionic) generators led to both globally supersymmetric and locally supersymmetric (supergravity) theories [1].  These models were devised with the purpose of having them be invariant under a supersymmetry that had been defined at the outset.  However, there is also the possibility of a model having both Bosonic and Fermionic symmetries whose algebra is not that of supergravity.  Such a model has been considered in ref. [2].

In applying the Henneaux-Teitelboim-Zanelli (HTZ) formalism [3] to supergravity in $2 + 1d$ [4] in order to find the symmetries (both Bosonic and Fermionic) that are present in the model, it becomes apparent that the same approach can be applied to find similar symmetries occuring in a $3 + 1$ dimensional model in which a gauge field with a topological action couples to a massless Majorana spin $3/2$ field.

In the next section we will introduce such a model and then analyze its symmetries using the HTZ formalism.  From the constraints present in the model, it can be quantized.  The conventions used appear in the appendix.

\section{The Model}

The Einstein-Cartan (EC) action for gravity in $3 + 1$ dimensions is 
\begin{equation}
S_{EC} = \int d^4 x\,ee^m_\mu e_\nu^n R_{mn}^{\mu\nu} \qquad (e = \det \,e_\mu^a)
\end{equation}
where $e_\mu^m$ is the vierbein and $R_{mn\mu\nu}$ is the field strength associated with the spin connection $w_{mn\mu}$,
\begin{align}
R_{mn\mu\nu} &= \partial_\mu w_{mn\nu} - \partial_\nu w_{mn\mu}\nonumber \\
& \hspace{1.2cm} + w^{\;\,k}_{m\mu} w_{kn\nu} - w^{\;\,k}_{m\nu} w_{kn\mu}.
\end{align}
This action can be rewritten as
\begin{equation}
S_{EC} = \frac{1}{4} \int d^4x \,\epsilon^{mnk\ell} \epsilon^{\mu\nu\lambda\sigma} e_{m\mu} e_{n\nu} R_{k\ell\lambda\sigma}.
\end{equation}

In this action, we shall treat $e_\mu^m$ and $w_{mn\mu}$ as being independent.  (One could take $w_{mn\mu}$ as being a function of $e_\mu^m$ that is found by solving the equation of motion for $w_{mn\mu}$.)

In coupling the gravitational field to matter fields, one employs the vierbein $e_\mu^m$.  However, if one wishes to deal with only the uncoupled gravitational field, then one could work with the fields
\begin{equation}
E_{\mu\nu}^{mn} = \frac{1}{4} \epsilon^{mnk\ell} \epsilon_{\mu\nu\lambda\sigma} e_k^\lambda e_\ell^\sigma
\end{equation}
so that the EC action takes the topological form
\begin{equation}
S_{EC} = \int d^4x E^{mn\mu\nu} R_{mn\mu\nu}.
\end{equation}
The EC action in $2 + 1$ dimensions when written in terms of the dreibein and spin connection is automatically in topological form
\begin{equation}
S_{EC} = \int d^3x \,\epsilon^{\mu\nu\lambda} e_\mu^i R_{\nu\lambda i}
\end{equation}
where now
\begin{equation}
R_{\mu\nu i} = \partial_\mu w_{\nu i} - \partial_\nu w_{\mu i} - \epsilon_{ijk} w_\mu^{\;j} w_\nu^{\;k} .
\end{equation}

In supergravity, one couples a Majorana spin $3/2$ field $\psi_\mu$ to the gravitational field.  In $3 + 1$ dimensions, this coupling is given by the action 
\begin{equation}
S_{3/2} = \int d^4x \epsilon^{\mu\nu\lambda\sigma} \overline{\psi}_\mu e_\nu^i \gamma_i \left( \partial_\lambda + \frac{i}{4} w_\lambda^{mn} \sigma_{mn}\right)\gamma_5 \psi_\sigma 
\end{equation}
while in $2 + 1$ dimensions we have
\begin{equation}
S_{3/2} = \int d^3x \,\epsilon^{\mu\nu\lambda} \overline{\psi}_\mu \left(\partial_\nu + \frac{i}{2} \gamma^i w_{\nu i}\right)\psi_\lambda.
\end{equation}
In ref. [4], a canonical analysis of supergravity in $2 + 1$ dimensions (as defined by eqs. (6,9)) is performed.  Repeating this exercise in $3 + 1$ dimensions is considerably more complicated.  (Just defining the appropriate Dirac Bracket (DB) [5] arising from the primary second class constraints which follow from the canonical momenta conjugate to $w_{mni}$ is prohibitively difficult. A canonical analysis of supergravity in $3 + 1$ dimensions in which a specific gauge is chosen at the outset is given in ref. [7].  See also refs. [8], [9].)  From the first class constraints in $2 + 1$ dimensional supergravity one can find a set of gauge transformations (both Bosonic and Fermionic) that leave the actions invariant; these transformations have a closed algebra and no auxiliary fields are required.  It would be a worthy achievement to obtain the same result for supergravity in $3 + 1$ (or indeed, $10 + 1$) dimensions.  However, as this enterprise is so difficult, we will instead consider a model in $3 + 1$ dimensions in which many of the features of supergravity in $2 + 1$ dimensions occur, making it possible to analyze its canonical structure using the Dirac constraint formalism [5].

In this model, we consider an $O(3)$ gauge field $A_\mu^a$ with a topological action that couples to a Majorana field $\psi_\mu^a$ with spin $3/2$.  The Lagrangian is given by
\begin{equation}
\mathcal{L} = - \frac{1}{2} \phi_{\mu\nu}^a F^{a\mu\nu} + \frac{1}{2} \epsilon^{\mu\nu\lambda\sigma} \overline{\psi}_\mu^{\,a} \gamma_\nu D_\lambda^{ab} \gamma_5 \psi_\sigma^b 
\end{equation}
where 
\begin{equation}
F_{\mu\nu}^a = \partial_\mu A_\nu^a - \partial_\nu A_\mu^a + \epsilon^{abc}A_\mu^b A_\nu^c
\end{equation}
and
\begin{equation}
D_\mu^{ab} = \partial_\mu \delta^{ab} + \epsilon^{apb} A_\mu^p .
\end{equation}
The field $\phi_{\mu\nu}^a$ in eq. (10) is a Lagrange multiplier field, much like the field $e^i_\mu$ in eq. (6). In the next section, we obtain the constraints associated with this model, and from the first class constraints obtain the symmetries that leave the action following from $\mathcal{L}$ in eq. (10) invariant.  These constraints are both Bosonic and Fermionic and satisfy a closed algebra that is distinct from the algebra of constraints in supergravity.

\section{Canonical Analysis}

The Lagrangian of eq. (10) can be rewritten
\begin{align}
\mathcal{L} &= - \frac{1}{2} \phi_{ij}^a F_{ij}^a + \phi_{0i}^a F_{0i}^a +
 \epsilon^{ijk}\bigg( \overline{\psi}_0^a \gamma_i D_j^{ab} \gamma_5 \psi_k^b \\
 & \hspace*{1cm} - \frac{1}{2} \overline{\psi}_i^{\,a} \gamma_0 D_j^{ab} \gamma_5 \psi_k^b + \frac{1}{2} \overline{\psi}_i^{\,a} \gamma_j D_0^{ab} \gamma_5 \psi_k^b\bigg).\nonumber
\end{align}
It is now apparent now that the momenta conjugate to $\phi_{ij}^a$, 
$\phi_{0i}^a$, $A_0^a$, $A_i^a$, $\psi_0$ and $\psi_i$ are respectively
\[
I\!\!P_{ij}^a = I\!\!P_i^a = p^a = 0 \eqno(14a-c)
\]
\[ p_i^a = \phi_{0i}^a \eqno(14d) \]
\[ \pi_0^a = 0 \eqno(14e) \]
\[ \hspace{2cm}\pi_i^a = -\frac{1}{2} \epsilon_{ijk} \overline{\psi}_j \gamma_k \gamma_5. \eqno(14f) \]
Eqs. (14b,d) constitute a set of Bosonic second class constraints
\[ \theta_{1i}^a = I\!\!P_i^a, \qquad \theta_{2i}^a = p_i^a - \phi_{0i}^a \eqno(15a,b) \]
such that we have the Poisson Bracket
\[\left\lbrace  \theta_{1i}^a,  \theta_{2j}^a \right\rbrace = \delta^{ab} \delta_{ij} .\eqno(16) \]
So also, the constraints of eq. (14f) are primary second class as well since
\[ \chi_i^a = \pi_i^a + \frac{1}{2} \epsilon_{ijk} \overline{\psi}_j \gamma_k \gamma_5 \eqno(17) \]
satisfies
\[ \left\lbrace \chi_i^{aT}, \chi_j^b \right\rbrace = -\epsilon_{ijk} C \gamma_k \gamma_5 .\eqno(18) \] 
From eqs. (16) and (18) we find the DB 
\[ \left\lbrace A,B \right\rbrace^* = \left\lbrace A,B \right\rbrace - 
\bigg[ \left\lbrace A,\theta_{2i}^a \right\rbrace  
\left\lbrace \theta_{1i}^a,B \right\rbrace - 
\left\lbrace A,\theta_{1i}^a \right\rbrace  
\left\lbrace \theta_{2i}^a,B \right\rbrace \nonumber \]
\[\hspace{1cm} + \left\lbrace A,\chi_i^a \right\rbrace  \left( \frac{i}{2}\gamma_j \gamma_i \gamma _0 C \right) 
\left\lbrace \chi_j^a,B \right\rbrace \bigg]. \eqno(19) \]
From this DB we see that
\[ \left\lbrace \psi_i^a, \overline{\psi}_j^b \right\rbrace^* = 
\frac{i}{2} \gamma_j \gamma_i \gamma_0 \delta^{ab} .\eqno(20) \]

The canonical Hamiltonian that follows from $\mathcal{L}$ in eq. (13) is 
\[\hspace{-4cm}\mathcal{H}_c = \dot{\psi}_i^{aT} \pi_i^{aT} + \dot{A}_i^{a} p_i^a - \mathcal{L}\nonumber \]
\[= \frac{1}{2} \phi_{ij}^a F_{ij}^a + \epsilon_{ijk} \bigg( \frac{1}{2} \overline{\psi}_i^a \gamma_0 D_j^{ab}\gamma_5 \psi_k^b \eqno(21) \]
\[\hspace{1cm} - \overline{\psi}_0^a \gamma_i D_j^{ab} \gamma_5 \psi_k^b \bigg) - A_0^a \bigg( D_i^{ab} p_i^b \nonumber \]
\[ - \frac{1}{2} \epsilon_{ijk} \epsilon^{abc} \overline{\psi}_i^b \gamma_j \gamma_5 \psi_k^c \bigg).\nonumber \]
The primary constraints that are not second class give rise to the following secondary constraints: 
\[\hspace{-3.7cm} \left\lbrace I\!\!P_{ij}^a, \mathcal{H}_c\right\rbrace^* = -\frac{1}{2} F_{ij}^a \equiv -\frac{1}{2} \Phi_{ij}^a \eqno(22) \]
\[  \left\lbrace p_{0}^a, \mathcal{H}_0\right\rbrace^* = \left( D_i^{ab} p_i^b -\frac{1}{2} \epsilon_{ijk} \epsilon^{abc} \overline{\psi}_i^b \gamma_j \gamma_5 \psi_k^c\right) \equiv \Phi^a \eqno(23) \]
\[ \hspace{.5cm} \left\lbrace \pi_{0}^a, \mathcal{H}_c\right\rbrace^* = -C \left( 
 \epsilon_{ijk} \gamma_i D_j^{ab} \gamma_5 \psi_k^b \right) = -C
 \Psi \equiv -C\Psi_c .\eqno(24) \]
Not all of the constraints $\Phi_{ij}^a$ are independent, as on account of the Bianchi identity they are related by
\[ \epsilon_{ijk} D_{\;i}^{ab} \Phi_{jk}^{\;b} = 0. \eqno(25) \]

It is now possible to establish the following DB algebra
\[ \left\lbrace \Phi^a, \Phi^b \right\rbrace^* = \epsilon^{abc} \Phi^c \eqno(26a) \]
\[ \hspace{1.5cm}\left\lbrace \Psi^a, \overline{\Psi}^b \right\rbrace^* = \frac{1}{2}   \epsilon^{ijk} \epsilon^{apb} \Phi^p_{ij} \gamma_k\gamma_5 \eqno(26b) \]
\[ \left\lbrace \Psi^a, \Phi^b \right\rbrace^* = \epsilon^{abc} \Psi^c \eqno(26c) \]
\[ \left\lbrace \Phi^a_{ij}, \Phi^b \right\rbrace = \epsilon^{abc} \Phi^a_{ij} \eqno(26d) \]
with all other DB involving $\Phi^a$, $\Phi_{ij}^a$ and $\Psi^a$ vanishing.

The Hamiltonian of eq. (21) can be rewritten as 
\[ \mathcal{H}_c = \frac{1}{2} \phi_{ij}^a \Phi_{ij}^a - \overline{\psi}_0^a \Psi^a - A_0^a \Phi^a + \mathcal{H}_x \eqno(27) \]
where the ``extra'' part of $\mathcal{H}_c$ is 
\[ \mathcal{H}_x = \frac{1}{2} \epsilon_{ijk} \overline{\psi}_i^a \gamma_0 D_j^{ab} \gamma_5 \psi_k^b .\eqno(28) \]
We find that 
\[ \left\lbrace \Psi^a, \int d^3x\, \mathcal{H}_x \right\rbrace^* = 
- \frac{1}{2} \epsilon^{abc}\epsilon_{ijk}\Phi_{ij}^b \gamma_0 \gamma_5 \psi_k^c \eqno(29a) \]
and
\[\hspace{1.2cm} \left\lbrace \Phi^a, \int d^3 \mathcal{H}_x \right\rbrace^* = 
\left\lbrace \Phi^a_{ij}, \int d^3 \mathcal{H}_x \right\rbrace^* = 0 \eqno(29b) \]
and so by eqs. (26) and (29) we find that $(I\!\!P_{ij}^a, p_0^a, \pi_0^a)$ are primary first class constraints, $(\Phi_{ij}^a, \Phi^a, \Psi^a)$ are the corresponding secondary first class constraints, and there are no further first class constraints.

These first class constraints lead to a generator of gauge transformations that is of the form
\[ G = a_{ij}^a I\!\!P_{ij}^a + b^a p_0^a + \pi_0^a c^a + \alpha_{ij}^a \Phi_{ij}^a + \beta^a \Phi^a + \overline{\gamma}^a \Psi^a .\eqno(28) \]
By HTZ, the coefficients of the secondary constraints $\alpha_{ij}^a$, $\beta^a$ and $\overline{\gamma}^a$ are taken to be field independent.  The coefficients of the primary constraints then follow from the HTZ equation   [3].  We find that
\[a_{ij}^a = 2D_0^{ab} \alpha_{ij}^b + \epsilon^{abc} \left[ \beta^b \phi_{ij}^c + \overline{\gamma}^b  \epsilon_{ijk} \left( \gamma_k \gamma_5 \psi_0^c + \gamma_0 \gamma_5 \psi_k^c\right) \right] = 0 \eqno(31a) \]
\[\hspace{-7.65cm} b^a = -D_0^{ab} \beta^b \eqno(31b) \]
\[\hspace{-5.9cm} \overline{c}^a = - D_0^{ab} \overline{\gamma}^b + \epsilon^{abc} \beta^b \overline{\psi}_0^c \eqno(31c) \]
so that 
\[ G = I\!\!P_{ij}^a \bigg[ 2D_0^{ab} \alpha_{ij}^b + \epsilon^{abc} \bigg( \beta^b \phi_{ij}^c + \epsilon_{ijk} \overline{\gamma}^b (\gamma_k \gamma_5 \psi_0^c \eqno(32) \]
\[ \hspace{2cm}+ \gamma_0 \gamma_5 \psi_k^c )\bigg)\bigg] - p_0^a D_0^{ab} \beta^b\nonumber \]
\[+ \pi_0^a\left( D_0^{ab} \gamma^b - \epsilon^{abc} \beta^b \psi_0^c   \right) \nonumber \]
\[\hspace{1.3cm} + \alpha_{ij}^a \Phi^a_{ij} + \beta^a \Phi^a + \overline{\gamma}^a \Psi^a .\nonumber \]
One can now compute the change in a dynamical variable $X$ generated by $G$, $\delta X = \left\lbrace X, G \right\rbrace^*$.  We find that 
\[ \hspace{-5cm}\delta A_\mu^a = D_\mu^{ab} \beta^b \eqno(33a) \]
\[\hspace{-2.7cm} \delta \psi_\mu^a = -D_\mu^{ab} \gamma^b - \epsilon^{abc}\psi_\mu^b \beta^c \eqno(33b)\]
\[ \delta \phi_{\mu\nu}^a = \epsilon_{\mu\nu\lambda\sigma} D^{ab\lambda} \theta^{b\sigma} - \epsilon_{\mu\nu\lambda\sigma} \epsilon^{abc} \overline{\psi}^{b\lambda} \gamma^\sigma \gamma_5 \gamma^c \eqno(33c) \]
where in eq. (33c), 
\[\alpha_{ij}^a = - \frac{1}{2} \epsilon_{ijk} \theta_k^a,\quad  \theta_0^a = 0.  \eqno(34)
\]

The gauge transformation associated with the gauge parameter $\beta^a$ is an ordinary non-Abelian gauge transformation.  The one associated with $\theta^a$ holds as a result of the Bianchi identity.  The Fermionic gauge symmetry associated with the gauge parameter $\gamma^a$ mixes the Bosonic field $\phi_{\mu\nu}^a$ with the Fermionic field $\psi_\mu^a$.  None of these transformations are associated with space-time symmetry and hence this is not a supergravity model.

Finally, if $G_I$ is the generator associated with the gauge functions ($\alpha_{I\,ij}^a$, $\beta_I^a$, $\gamma_I^a$), then by computing $\left\lbrace G_I, G_J \right\rbrace$* we see that the DB of the generators $G_I$ and $G_J$ gives rise to a generator $G_K$ with 
\[ \alpha_{Kij}^a = -\frac{1}{2} \epsilon_{ijk}\epsilon^{abc} \overline{\gamma}_I^b \gamma_k \gamma_5 \gamma_J^c + \epsilon^{abc} \left( \beta_I^b \alpha_{Jij}^c + \alpha_{Iij}^b \beta_J^c\right) \eqno(35a) \]
\[\hspace{-6cm} \beta_K^a = \epsilon^{abc} \beta_I^b \beta_J^c \eqno(35b)\]
\[ \hspace{-4cm}\overline{\gamma}_K^a = \epsilon^{abc} \left( \beta_I^b \overline{\gamma}_J^a - \beta_J^b \overline{\gamma}_I^a\right). \eqno(35c) \]
We thus have a closed algebra for the Bosonic and Fermionic gauge transformations that leaves the action of eq. (10) invariant.  No auxiliary fields are required.

In this model, there are 20 Bosonic degrees freedom in phase space ($A_\mu^a$, $\phi_{\mu\nu}^a$ and their conjugate momenta) plus 16 Fermionic degrees of freedom ($\psi_\mu^a$ and its conjugate momentum).  From eqs. (14b,d) we have six Bosonic second class constraints and from eq. (14f), six Fermionic second class constraints.  By eqs. (14a,c) there are four primary first class Bosonic constraints; in addition there are three secondary first class Bosonic constraints (by eqs. (22, 23, 25)).  The four  first class Fermionic constraints are given by eqs. (14e) (primary) and (24) (secondary).  When one accompanies each first class constraint with an appropriate gauge condition, we see that of the 36 degrees of freedom present, only two are left in phase space on account of the 34 constraints present.  This single degree of freedom is Fermionic.

\section{Discussion}

The Dirac constraint formalism [5], when accompanied by the HTZ procedure for deriving the generator of gauge transformations from the first class constraints in a theory [3], has proven to be a useful way of analyzing both Bosonic and Fermionic gauge symmetries.  This has been demonstrated both by considering the spinning particle [6], and supergravity in $2 + 1$ dimensions [4].  Although it has not yet been feasible to apply this analysis to supergravity in $3 + 1$ dimensions, we have shown it to be possible to adapt the treatment of supergravity in $2 + 1$ dimensions to formulate a $3 + 1$ dimensional model in which the Bosonic and Fermionic gauge transformations form a closed algebra that is not that of supergravity.  Quantization can proceed in the way outlined in ref. [4]; this will result in both Bosonic and Fermionic ``ghosts''.

\section*{Acknowledgements}

Roger Macleod had useful suggestions.

\section*{Appendix - Conventions and Notation}

We employ the metric $g^{\mu\nu} = \mathrm{diag} (+ - - -)$ with $\epsilon^{0123} = +1 = \epsilon^{123} = \epsilon_{123}$.
For Dirac matrices,
\begin{equation}
\gamma^0 = \gamma_0 = \left( \begin{array}{cc}
0 & 1 \\
1 & 0 \end{array}\right) \quad 
\gamma^i = -\gamma_i =  \left( \begin{array}{cc}
0 & -\sigma_i \\
\sigma_i & 0 \end{array}\right) \quad
\sigma_{\mu\nu} = \frac{1}{4i}\left[ \gamma_\mu , \gamma_\nu \right] \nonumber
\end{equation}
where $\sigma_i$ is a Pauli spin matrix.  If $\gamma^5 = \gamma_5 = i\gamma^0 \gamma^1 \gamma^2 \gamma^3$ then 
\begin{equation}
\gamma^\mu\gamma^\nu\gamma^\lambda = \gamma^\mu g^{\nu\lambda} - \gamma^\nu g^{\mu\lambda} + \gamma^\lambda g^{\mu\nu} + i \epsilon^{\mu\nu\lambda\rho} \gamma_\rho \gamma^5 \nonumber
\end{equation}
and
\begin{equation}
\gamma_i\gamma_j = -\delta_{ij} - i\epsilon_{ijk}\gamma_k\gamma_0\gamma_5, \quad \epsilon_{ijk} \gamma_j\gamma_\ell\gamma_k = -2i\delta_{i\ell} \gamma_0\gamma_5. \nonumber
\end{equation}
The ``charge conjugation'' matrix $C = -C^T = C^{-1} = \gamma_0\gamma_2$ satisfies $C \gamma_\mu C^{-1} = -\gamma_\mu^T$. Furthermore we have 
$\gamma_0\gamma_\mu \gamma_0 = \gamma_\mu^\dagger$.\

A spinor $\psi$ is Majorana if $\psi = \psi_c \equiv C\overline{\psi}^T$ where $\overline{\psi} = \psi^\dagger \gamma_0$. This implies that $\overline{\psi} = - \psi^TC$.  If $\chi$ is also a Majorana spinor, then $\overline{\psi} \chi = (\overline{\psi} \chi)^\dagger = \overline{\chi}\psi$ with both $\chi$ and $\psi$ being Grassmann.

We use the left derivative for Grassmann variables $\theta_a$ so that 
\begin{equation}
\frac{d}{d\theta_a} (\theta_b \theta_c) = \delta_{ab} \theta_c - \delta_{ac}\theta_b, \quad \frac{d}{dt} F(\theta(t)) = \dot{\theta}(t) F^\prime(\theta(t)). \nonumber 
\end{equation}
If $(q_i,  p_i = \frac{\partial L}{\partial \dot{q}_i})$ and 
$(\psi_i,  \pi_i = \frac{\partial L}{\partial \dot{\psi}_i})$ are Bosonic and Fermionic canonical variables respectively, then for Poisson Brackets we have 
\begin{align}
\left\lbrace B_1, B_2 \right\rbrace &= \left(B_{1,q} B_{2,p} - B_{2,q} B_{1,p} \right) + \left(B_{1,\psi} B_{2,\pi} - B_{2,\psi} B_{1,\pi} \right) \nonumber \\
\left\lbrace B, F \right\rbrace &= \left(B_{,q} F_{,p} - F_{,q} B_{,p} \right) + \left(B_{,\psi} F_{,\pi} + F_{,\psi} B_{,\pi} \right) = - \left\lbrace F, B \right\rbrace \nonumber \\
\left\lbrace F_1, F_2 \right\rbrace &= \left(F_{1,q} F_{2,p} + F_{2,q} F_{1,p} \right) - \left(F_{1,\psi} F_{2,\pi} + F_{2,\psi} F_{1,\pi} \right).  \nonumber
\end{align}
The Hamiltonian is
\begin{equation*}
H = \dot{q}_i p_i + \dot{\psi}_i \pi_i - L. 
\end{equation*}

\end{document}